\documentstyle[12pt,epsfig]{article}
\begin{document}
\begin{titlepage}
\begin{center}
\vspace*{3cm}

{
\Large  \bf Charge fluctuations in a final state with QGP
}
\vspace{2cm}

\begin{author}
\Large
K. Fia{\l}kowski\footnote{e-mail address: uffialko@thrisc.if.uj.edu.pl},
R. Wit\footnote{e-mail address: wit@thrisc.if.uj.edu.pl}

\end{author}

\vspace{1cm}

{\sl M. Smoluchowski Institute of Physics\\ Jagellonian University \\

30-059 Krak{\'o}w, ul.Reymonta 4, Poland}

\vspace{3cm}

\begin{abstract}
Charge fluctuations as a  possible signal of 
quark - gluon plasma (QGP) were recently suggested. A short summary of comments
presented on this subject is given and supplemented by a discussion of the coexistence of
pions produced "directly" and through a QGP phase. Such a coexistence may obscure
the expected plasma signal similarly to the  effects considered in the comments 
mentioned above.
\end{abstract}

\end{center}
\vspace{1cm}



\vspace{1cm}

\noindent

 20 January, 2001 \\

\end{titlepage}

\par
It was suggested recently \cite{Sh1,Sh2} that the "primordial 
frozen fluctuations"
may be used to signal the formation of a quark-gluon plasma (QGP) in the early 
stage of high energy interaction of heavy ions. Similar considerations led
to more detailed predictions for measurable quantities \cite{JK,AHM}. In   
particular, Jeon and Koch \cite{JK} argued that the event-by-event 
fluctuations of charge (or, equivalently, of the ratio of positive to 
negative pions) in a restricted rapidity range may be used as a tool to signal the 
possible formation of quark - gluon plasma (QGP). They compared the ratio of charge 
dispersion squared to charged multiplicity in two models: the "pion gas" and the QGP 
and conclude that they differ by a factor of five. Similar result were found for the 
dispersion of "positive-to-negative" ratio for pions. The dramatic difference 
may be easily  understood as the reflection of small quark charges as compared to 
hadrons and of zero gluon charges.
\par
Jeon and Koch concluded that the strong decrease of charge fluctuations in the QGP as 
compared to a pion gas should be seen as an "unmistakable signal of QGP 
formation from 
'Day-1' measurements" at RHIC. They added some caveat about resonances and other 
correlation effects which may reduce the fluctuations in the "pion phase", but these 
were claimed to be minor corrections.
\par
After appearance of this paper we indicated \cite{FW}  that
the values of  the ratio of charge dispersion squared to charged multiplicity $R$ quoted
in Ref.\cite{JK} are unrealistic. In any reasonable model of standard (non-QGP)
hadron production these values depend strongly on the rapidity bin width and already 
for the values of a few units they are as low as those quoted for QGP. This was 
demonstrated on the example of the JETSET/PYTHIA generator \cite{SJO}. This dependence
stems mainly from the effect of global charge conservation, which was discussed
simultaneously elsewhere \cite{HJ}.
\par
This prompted the authors of Ref.\cite{JK} to introduce corrections in the journal version
of their paper \cite{JK2} and to write an extended paper, in which the r\^ole of the
corrections for global charge conservation is discussed in more detail \cite{BJK}.
It was suggested that after dividing the measured ratio $R$ by a correction factor $1-p$,
one recovers the originally suggested values for "QGP" and "non-QGP" cases.
{\renewcommand{\thefootnote}{\fnsymbol{footnote}}
\footnote{By $p$ we denote here the probability that a produced pion falls into the considered 
rapidity bin. Obviously, $p$ is an increasing function of bin width.}}
Thus the authors still regard their analysis as providing a clear signal of possible 
QGP formation.
\par
This claim was questioned in a few papers. It was pointed out that the predictions of
string models of "non-QGP" hadroproduction for the $R$ ratio 
differ quite strongly from those of thermodynamical models \cite{BR}. This makes 
the possible distinction of "QGP case" rather questionable. Also the results from
different MC generators were analyzed in detail \cite{GTZ}. It was argued that
the possible independence of $R$ on the impact parameter may suggest that it
measures a QCD scale rather than QGP effects. Another effect neglected
in Refs.\cite{JK2,BJK} is the final state rescattering, which may reduce
significantly the difference between the "QGP" and "non-QGP" values of $R$ \cite{SS}.
\par
In this note we consider one more obvious effect, which may impede the observation
of a QGP signal: the coexistence of hadrons from two sources in the same event. Indeed,
it seems rather unrealistic to assume that all the pions observed in the event come
from the hadronization of QGP. Even for the most central heavy ion collisions
some hadrons are likely to be produced in the collisions of "peripheral" nucleons, 
which may be described by string- or thermodynamical models. In other words, it
seems unlikely that the "QGP bubble" covers all the interaction volume.
\par
Let us consider a toy model in which there is a common multiplicity distribution 
${\it P(N)}$ for all pions produced in full phase space. Let us assume that there
are binomial probability distributions for this pion to be a "direct" positive or
negative pion or an "effective quark" which later hadronizes in a pion. We assign
charges $\pm 1/3$ to these quarks (which is a rough average for realistic quarks and gluons) 
and impose global charge conservation for produced
particles (neglecting thus fragmentation of initial nucleons, which does not affect
the spectra in the central rapidity region). We find then
\begin{equation}
P(N_+,N_-,K_+,K_-)=\sum_N {\it P(N)}B_q(N,N_+)B_q(N,N_-)\delta_{N,N_++N_-+K_++K_-}   
\delta_{N_+-N_-,(K_--K_+)/3}
\end{equation}
where $N_+$, $N_-$ denote "direct" positive and negative pions, $K_+,~K_-$ positive
and negative quarks (which  hadronize later into pions) and $B_q(N,M)$ is the binomial
probability distribution given by
\begin{equation}
B_q(N,M)=\frac{N!}{M!(N-M)!}q^M(1-q)^{N-M}.
\end{equation}
\par
One should add that we assume implicitly that each "effective quark" hadronizes into
one charged hadron. In the old-fashioned recombination model, in which there is one
 charged hadron per $2/3$ of
$q\overline q$ {\it pairs}  the suppression effect from QGP phase would be much weaker. 
\par
Now let us consider the rapidity interval of the width $\Delta$. Assume that for each 
type of particles the probability to fall into this interval is given by the same
function of bin width $\Delta$ (denoted by $p$, as before) and  
that the probability to find a given number of each type of particles is also given
by a binomial formula. Then
 \begin{equation}
p(n_+,n_-,k_+,k_-) = \sum P(N_+,N_-,K_+,K_-)B_p(N_+,n_+)B_p(N_-,n_-)B_p(K_+,k_+)B_p(K_-,k_-).   
\end{equation}
This allows us to calculate the generating function of charge distribution in the 
rapidity interval 
\begin{equation}
G_Q^{\Delta}(z) \equiv \sum_Q z^Q {\cal P}(Q) = \sum z^{n_+-n_-+k_+/3-k_-/3}p(n_+,n_-,k_+,k_-).
\end{equation}
We get
\begin{eqnarray}
& &G_Q^{\Delta}(z) = 
{\cal G}[(1-p+z^{1/3}p)^{1/2}(1-p+z^{-1/3}p)^{1/2}\ast\\ 
& &\ast (1-q+\frac{q(1-p+zp)(1-p+z^{-1/3}p)}{(1-p+z^{1/3}p)^2}) 
(1-q+\frac{q(1-p+p/z)(1-p+z^{1/3}p)}{(1-p+z^{-1/3}p)^2})]\nonumber
\end{eqnarray}
where ${\cal G}[y]$ is the generating function of the multiplicity distribution
in the entire phase space
\begin{equation}
{\cal G}[y] \equiv \sum_N y^N {\it P(N)}.
\end{equation}
\par
The derivatives of the generating function taken at $1$ are
the distribution moments. In particular
\begin{equation}
\frac{d{\cal G}}{dy}|_{y=1} \equiv \sum_N N {\it P(N)} \equiv <N>.
\end{equation}
For the moments of the charge distribution we find then
\begin{eqnarray}
<Q> &\equiv& \frac{dG}{dz}|_{z=1} =0,\\ 
<Q^2>&=&<Q(Q-1)> \equiv \frac {d^2 G}{dz^2}|_{z=1}=<N>\frac{16q+1}{9}p(1-p).
\end{eqnarray} 
Thus the ratio of charge dispersion squared to average charged multiplicity is
\begin{equation}
R(\Delta) = \frac{D^2_Q}{<n_{ch}>} = \frac{16q+1}{9}[1-p(\Delta)].
\end{equation}
\par
Obviously, for $q=1/2$ (which corresponds to all pions produced directly) we recover
the standard result $D^2_Q~/<n_{ch}>~=1-p$, and for $q=0$ (when all pions come from 
plasma) we get a suppression factor of $1/9$
\footnote[1]{This very small value is a direct 
consequence of our assumptions for "effective quark" charges. In a model taking into 
account "true" quarks and gluons this factor is usually $1/5$ to $1/4$ \cite{JK2}.}.
 For the realistic values of $2q$ around
$0.5$ (which corresponds to a similar number of pions from both sources) the 
suppression factor is above $1/2$ instead of $1/9$.
\par 
Our toy model represents quite well the broad class of models in which the final 
state is a superposition of pions from two independent sources. In fact, the results 
hardly change if we impose charge conservation separately for both sources. Therefore 
we may conclude that the presence of pions which do not come from plasma brings the
suppression factor much closer to $1$ than expected for QGP. If the fraction $r$ of 
pions comes "directly", and the 
suppression factor for plasma is $s$, we get approximately
\begin{equation}
R(\Delta) = \frac{D^2_Q}{<n_{ch}>} = [r+(1-r)s] [1-p(\Delta)].
\end{equation}
\par
As one sees, we get the same $1-p$ factor for both contributions from "non-QGP"
and "QGP", as suggested in Ref.\cite{BJK}. This is, however, because we assumed the same 
dependence of $p$ on $\Delta$ for "QGP" and "non-QGP" contribution, which is by no 
means obvious. In fact, it is more likely that the pions from QGP are more peaked 
at mid-rapidity. Then $q$ and $r$ increase with $\Delta$, which may cancel the decrease
due to the $1-p$ factor. This suggests that the change of the dependence of $R$ on 
$\Delta$ may be a similarly (in)significant signal of the onset of QGP as the 
drop in the value of $R$.
\par
Summarizing, we have shown that the coexistence of the "direct" pions and pions 
coming from plasma reduces the suppression effect expected from "pure QGP". 
Adding this reduction to other effects considered before (resonances and string effects 
reducing the $R$ value for "non-QGP", final state rescattering enhancing the 
value for QGP etc.) one may conclude that it is unlikely to find a 
strong suppression signal when the threshold for QGP formation is passed.
\\
\\

{\bf Acknowledgements}
\par
We would like to thank Kacper Zalewski for reading the manuscript and helpful remarks.
 One of us (RW) is grateful for a partial financial support by the KBN grant \# 2 
P03B 019 17.


\begin{thebibliography}{99}
\bibitem{Sh1}E.V. Shuryak, {\it Phys. Lett.} {\bf B423}, 9 (1998).
\bibitem{Sh2}E.V. Shuryak, {\it Nucl. Phys.} {\bf A661}, 119c (1999).
\bibitem{JK}S. Jeon and V. Koch, {\it Charged Particle Ratio Fluctuation as a 
Signal for QGP},\\ hep-ph/0003168 v2 (2000).
\bibitem{AHM} M. Asakawa, U. Heinz and B. M\"uller, {\it Phys. Rev. Lett.} {\bf 85}, 2072 (2000).
\bibitem{FW}K. Fia{\l}kowski and R. Wit, e-print, {\it Are charge fluctuations
a good signal for QGP?}, hep-ph/0006023 (2000).
\bibitem{SJO}T. Sj\"ostrand and M. Bengtsson, {\it Comp. Phys. Comm.}
{\bf 43}, 367 (1987); T. Sj\"ostrand, {\it Comp. Phys. Comm.} {\bf 82}, 74 
(1994).
\bibitem{HJ}H. Heiselberg and A.D. Jackson, {\it Anomalous Multplicity Fluctuations from 
Phase Transitions in Heavy Ion Collisions}, e-print nucl-th/0006021 v2 (2000).
\bibitem{JK2}S. Jeon and V. Koch, {\it Phys. Rev. Lett.} {\bf 85}, 2076 (2000).
\bibitem{BJK}M. Bleicher, S. Jeon and V. Koch, {\it Phys. Rev.}
{\bf C62}, 061902 (2000).
\bibitem{BR}F.W. Bopp and J. Ranft, {\it Charged Particle Fluctuations as a 
Signal of the Dynamics of Heavy Ion Processes}, e-print hep-ph/0009327.
\bibitem{GTZ}C. Gale, V. Topor Pop and Q.H. Zhang, {\it Is Charged Particle Ratio
Fluctuation a Signature of QGP?}, (2001).
\bibitem{SS}E.V. Shuryak and M.A. Stepanov, {\it When can long-range charge 
fluctuations serve as a QGP signal?}, e-print hep-ph/0010100 (2000).

\end{thebibliography}
\end{document}